# Geophysical Electromagnetics: A retrospective, DISC 2017, and a look forward

Douglas W. Oldenburg[1], Lindsey J. Heagy[1,2] & Seogi Kang[1,3]

[1] University of British Columbia Geophysical Inversion Facility (doug@eoas.ubc.ca)
[2] University of California Berkeley, Department of Statistics (lheagy@berkeley.edu)
[3] Stanford University, Department of Geophysics (sgkang09@stanford.edu)

## Abstract

Geophysical electromagnetics (EM) plays an important role in mineral exploration and is increasingly being used to help solve other problems of relevance to society. In this article we reflect, from our perspective at University of British Columbia (UBC), on the development of EM geophysics over the years and on our attempts to enhance its understanding, and its visibility and usefulness to the community. The availability of open-source resources, and a shift within the EM community towards collaborative practices for sharing and creating software and educational resources, has been a driver of progress towards these goals. In this article we provide some background about this trajectory and how the SEG Distinguished Instructor Short Course (DISC) was a catalyst in our development of software and resources and in our broader goal of creating more collaborative connections within the EM community.

## Introduction

We begin with some historical background about development of applied geophysics at University of British Columbia (UBC) and the Geophysical Inversion Facility (GIF). The UBC-GIF was established in 1989 with a mandate to serve as an interface between academic research and industrial applications. At an inaugural open house we invited geoscientists from industry, academia and government to participate in presentations and discussion about the potential for inversion to enhance the information they were obtaining from geophysical data. Figure 1 shows the group associated with this event; of particular note is the presence of Misac Nabighian who is an iconic figure in electromagnetic geophysics.

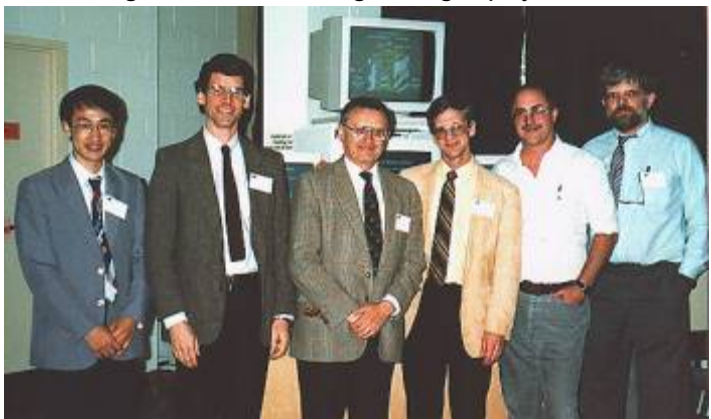

Figure 1: Founding members of the UBC-GIF at the first UBC-GIF Open House in 1989. Left to right: Yaoguo Li, Robert G. Ellis, Misac Nabighian, Doug Oldenburg, Rob Ellis and Greg Shore (https://www.eoas.ubc.ca/ubcgif/background/history.html).

Since then, the primary focus for GIF research has been on mineral exploration problems and, in particular, on the development of methodologies and software to invert various types of geophysical data. We began with 2D direct current (DC) resistivity and induced polarization (IP), followed up with 3D potential fields, 1D electromagnetics (EM), and progressively tackled the computationally challenging problems of inverting 3D EM data with many sources (Oldenburg and Li, 1994; Li and Oldenburg, 1996; Farquharson et al., 2003; Haber et al., 2007; Oldenburg et al., 2012). Our progress has paralleled advances in computational power, numerical solvers, and solving optimization problems. The other pillar of our success has been the steadfast support of mining companies who have motivated us by supplying problems and data, applied our software solutions, and provided feedback about needed improvements or functionality. The financial stability of our research program has resulted from the long-term commitment of the industrial sponsors. Seven[1] companies in our current consortium have continually sponsored our research for the last 30 years, although their names have changed because of mergers or acquisitions.

In addition to research, the GIF Outreach program was designed to disseminate information about the fundamentals of inversion and how to apply it to field data. The intended audiences have been expert and non-expert geophysicists, as well as engineers and geologists. In our formative years, we provided free workshops in DC/IP and magnetic inversions, and had booths at Canadian mining conferences. By reaching a diverse audience, and in particular non-geophysicists who would be involved in the decision making process of whether to use a geophysical technique, we felt we could increase the use and usefulness of geophysics. As a product of this work we developed, in the mid-1990's, the IAG (Inversion for Applied Geophysics) educational resource. This was a digital textbook which was distributed via a cd-rom and was also licensed and freely downloadable from UBC. It was meant to help inform non-experts (e.g. geologists and engineers) about the physical basis of geophysical surveys and fundamentals of inversion; it provided software targeted at educational purposes, and case histories regarding application. The IAG materials formed a foundation for subsequent resources, including the GPG (Geophysics for Practicing Geoscientists; https://gpg.geosci.xyz) which was developed to be the primary "textbook" for an applied geophysics course at UBC taken by geological engineers and geologists. More recently, it has been adopted as a course "textbook" by instructors at at least 5 other universities, and has been used by 31,000 people in the past year (Sept 2019-Sept 2020). Though many of the goals of the GIF Outreach vision have remained the same, advances in the web for sharing information and enabling collaboration over the past decade allow us to achieve these goals in ways that are far more effective.

With this background, it is understandable why we were enthusiastic when asked by the SEG to present the 2017 Distinguished Instructor Short Course (DISC). The founding philosophy of the DISC, as put forth by Peter Duncan, is "to be the crowning jewel in the SEG's CE [Continuing Education] program, uniting a world-class instructor with a leading-edge topic to create an educational event of global proportions." The DISC works in the following manner:

---

[1] The current members are: Barrick, Glencore, BHP, Vale, Teck, Anglo American, Rio Tinto. The original founding members in 1990 were: BHP, CRA Exploration, Cominco, Falconbridge, Hudson Bay Exploration and Development, INCO, Kennecott, Newmont, Noranda, Placer Dome, WMC with matching funds provided by NSERC.

an instructor and topic (heretofore related to seismic and hydrocarbons) is selected; the instructor puts together material for presentation and also writes a book which serves as a resource for those attending and which is also later sold through the SEG; the 1-day course is taken around the world and local hosts are required to organize the event at their location and generate enough participation so that registration fees cover the presentation costs. The DISC has been very successful in the past, often reaching more than two thousand participants in ~20+ countries.

When asked by the SEG to generate a course on a non-seismic topic, we saw this as an opportunity to make an impact, and increase the usefulness of EM in a wide variety of applications that span hydrocarbons, minerals, geothermal energy, groundwater, natural hazards, environmental and geotechnical problems.

### *Why a course on EM?*
The physical properties associated with EM are electrical conductivity/resistivity, magnetic permeability (often cast in terms of magnetic susceptibility), and electrical permittivity (cast as dielectric constant). These properties can play a valuable role in helping solve many practical problems. However, EM is not used to its full potential and, in many cases, has been oversold (Constable, 2010) or misused (Hodges, 2005). Part of this comes about through lack of understanding and incorrectly conceptualizing EM phenomena. In seismic surveys it is intuitive to visualize how a wave packet propagates through the medium, reflects/refracts at interfaces, and returns to the surface as a wavelet. EM is unintuitive, especially when working in the frequency domain where the signals are partitioned into "real" and "imaginary" parts. Historically, many EM techniques have been oversold with respect to their resolving power and their ability to get detailed information using just a few sources and receivers. For a 3D earth, just as in seismology, the survey must have 3D acquisition so that the entire volume under investigation is illuminated. This requires many sources and receivers which measure multiple components of the vector fields. We need lots of high quality data; there is no free-lunch! A last important point is that for almost all problems the geologic structure is 3D and there is topography. If the data are inverted using 1D or 2D assumptions there can be artefacts which, if interpreted geologically, might be quite wrong. Historically this has happened far too often and produces a net consequence that the EM geophysical technique "does not work", or that EM doesn't work. Period!

The good news is that there have been many achievements over the last decade and we now have a "perfect storm" for advancing the application of EM. The components are: (1) there are many applications where EM can play a role; (2) advances in instrumentation mean that large amounts of high quality data can be collected; (3) advances in computer hardware, HPC/cloud computing and computational software (e.g. linear solvers) mean that we can simulate, and invert in 3D, almost all types of EM data; (4) advances in web-tools allow us to communicate and promote collaboration; (5) there is a new cadre of brilliant young scientists who want to use EM to solve important societal problems. Our challenge was to design a course that consolidated these advances and presented them to a diverse audience world-wide and make a long-term impact. Thus we needed to determine our target audience, decide on material and how to present it.

### *Who is the audience?*
The diversity of applied problems we face as a society is immense and ranges from: (a)

resource exploration, production and monitoring (for minerals, hydrocarbons, water); (b) energy (geothermal power); (c) environmental (contaminants, and disposal or containment of nuclear wastes); (d) geotechnical (tunnels, infrastructure); (e) monitoring and evaluating effects of climate change. Some of this diversity is provided in Figure 2. Solving these problems requires a multidisciplinary team that could include geophysicists with a variety of expertise and backgrounds, as well as geologists, engineers, and other stakeholders. EM geophysics might play an important role, but it is rare that knowing the 3D distribution of one of the EM physical properties (e.g. resistivity), provides a unique trajectory towards a solution. The information from EM is only one piece of data that needs to be integrated into a holistic perspective of the problem. Since the DISC was to be presented in many countries, each with its own geoscience priorities and capabilities, we anticipated the need to provide material what would be of use to geophysicists who were well versed in EM, geophysicists who might be expert in other areas of geophysics (such as seismology), and other geoscientists from academia, industry or government. To make matters more complicated, some of the heavy users of EM were specialized in certain areas (e.g. airborne EM) but were unfamiliar with other techniques (e.g. grounded source, MT or GPR). There was also a bifurcation between those who used time domain (TDEM) and those who used frequency domain (FDEM). In summary, we expected a very diverse group!

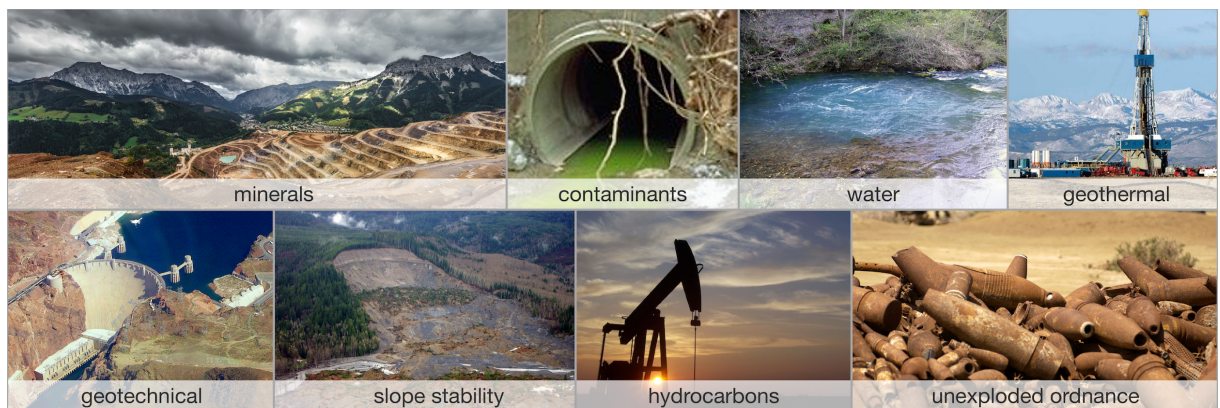

Figure 2. A sampling of applications where electromagnetic methods can play a role

In a multi-disciplinary applied problem, it is essential that all members of the team can communicate effectively; this means understanding the lexicon and the physical principles of EM, and also having tools to communicate. Here is where having web-based tools for collaboration (e.g. GitHub) and for sharing interactive educational resources, such as Jupyter Notebooks, play an essential role. These tools allow people to work together, share knowledge and results, and be more effective in solving the problem. The associated practices of collaborative development, peer review of work, and open sharing of new developments also promotes the building of a community. These are concepts that motivated our approach to the 2017 DISC and we wanted to share our vision with the community. In the sections below we talk about these items.

*The textbook*

A main component of previous DISC presentations was the physical textbook that accompanied the course. Because of the breadth of our course material, and our final objectives, we chose a different route. Our goal was to build an online, open-source, collaboratively developed resource, EM GeoSci (https://em.geosci.xyz). Its main components would be a mathematical foundation for EM (modelled closely by the classic chapter by Ward and Hohmann (1988)), a conduit for working with interactive simulations, a synopsis for the various types of EM surveys, and Case Histories which capture field examples of the use of EM to solve an applied problem. All of the content is licensed under Creative Commons, accessed through any browser, stored on GitHub, and anyone can contribute. This was a massive undertaking for the GIF group and for those geoscientists outside of UBC who contributed case histories. The resource is designed to be used either from the "bottom-up" [physics → surveys → case history applications] or "top-down" [case history application → relevant survey → underlying physics]. The resource is also a conduit to the apps which facilitate the interactive exploration of the physics of EM and illustrate inversion for obtaining a subsurface image from EM data.

*What are apps?*
The apps are self-contained Jupyter notebooks (Pérez & Granger, 2007) that allow the user to investigate fundamentals, ask questions, and interact with the physics of EM. Being able to plot charge densities, currents, electric and magnetic fields as a function of time, frequency, or space allows the user to gain insight and to self-test about his/her understanding. An app makes the equations presented in standard textbooks come alive. The apps can be run on the cloud through services such as Binder ([mybinder.org](mybinder.org); Project Jupyter et al., 2018) or they can be downloaded and run on a local machine. At the closure of DISC 2017 we had 26 apps that spanned DC resistivity, frequency and time domain, and natural sources. Figure 3 is an app for the 3-layer earth model. The user can explore the effects of changing electrode geometries, resistivities of the layers, and plotting total or secondary fields, currents, and charges.

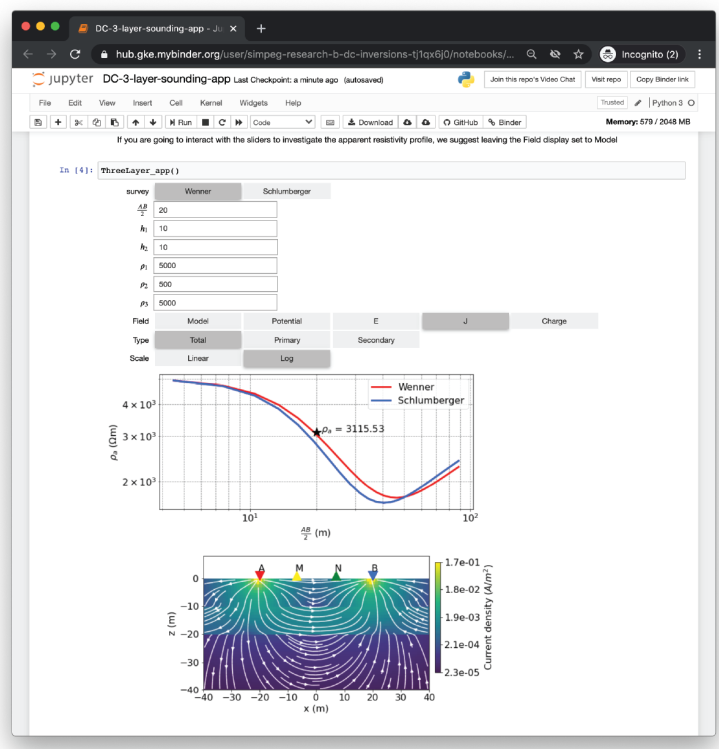

Figure 3. DC 3-layer app allows the user to obtain insight about the relationships of electrode geometry, resistivity of the earth, the currents and apparent resistivity value.

*What is a Case History?*

A case history demonstrates the use of the EM method to solve a problem. Our definition of case history is the same as what is standardly presented, but we have explicitly divided it into a Seven-Step procedure: (1) Problem, (2) Physical Properties, (3) Survey, (4) Data, (5) Processing, (6) Interpretation, (7) Synthesis. By parsing a case history into these segments, it makes a succinct document that facilitates impact and communication. The case histories for EM GeoSci.xyz were usually based upon published papers and links to those formative documents are provided. Each is a synopsis of the complete work and thus is valuable for new-comers to the field who want a quick overview of how an EM survey was beneficial, or show to management how a survey has been successful.

Figure 4 is an image of the EM GeoSci.xyz page for the case history concerning the inversion of DC and IP at Mt. Isa. This was one of the first 3D inversions of IP data carried out and in this case history the reader is taken through the various steps of survey design, processing, inversion and interpretation (Rutely et al., 2001). Throughout the case history there are links to other sections of EM GeoSci.xyz that provide background about the physics, field survey and equipment and general inversion procedures. The case history also compares the results obtained from present-day 3D inversion codes to those from two decades ago.

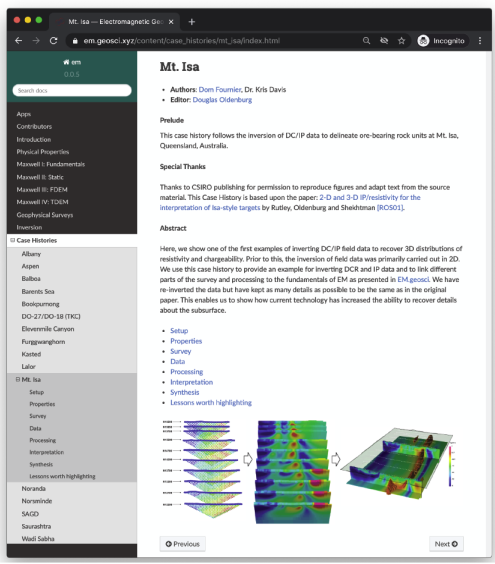

Seven steps for a case history:

1. Setup
2. Physical Properties
3. Survey
4. Data
5. Processing
6. Interpretation
7. Synthesis

Figure 4. Case histories on EM GeoSci.xyz follow a seven step framework. The image on the left is the page for the Mt. Isa case history.

# What was DISC 2017?

A primary goal for the DISC was to inform people about different EM surveys and how they might successfully be applied. We also wanted to use this opportunity to find what EM techniques were being used, and on which problems, in various locations around the world. We wanted to provide an opportunity for local groups to present their challenges and accomplishments, have a conversation, and open up channels for (international) collaboration. To accomplish this we turned the DISC into a two-day event. The first day was devoted to the DISC course, a primarily lecture-based day; the second day was DISC Lab, our addition of a day targeted at conversations where participants presented their work and interests in EM.

## DISC Course

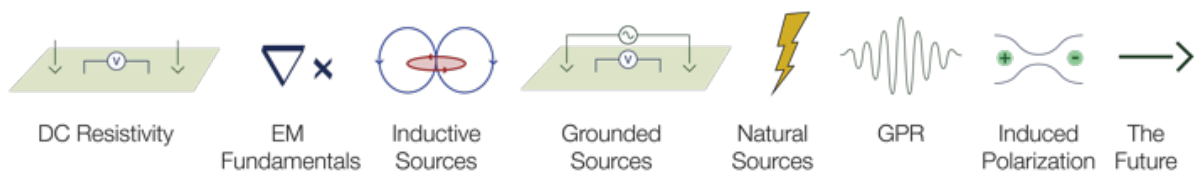

Figure 5. Eight subject elements that made up the DISC course.

Our course was divided into segments, as shown in Figure 5. For each segment we presented the fundamental physics, introduced apps for running numerical simulations and exploring the physics, provided an overview of the applicable EM survey method, and concluded with a case history. It was a lot of material (both for the listeners and presenter) but it provided, for a first time for many people, a complete spectrum of EM techniques. DC resistivity, with its understanding in terms of charge buildup, formed a foundation which was useful throughout the course. In EM fundamentals, the apps helped greatly in understanding response functions in the frequency and time domain, and clarified the (often confusing) aspect of frequency domain signals that get partitioned into real and imaginary parts. Inductive sources, which are the basis for all airborne systems, were next on the list. (See Figure 4)

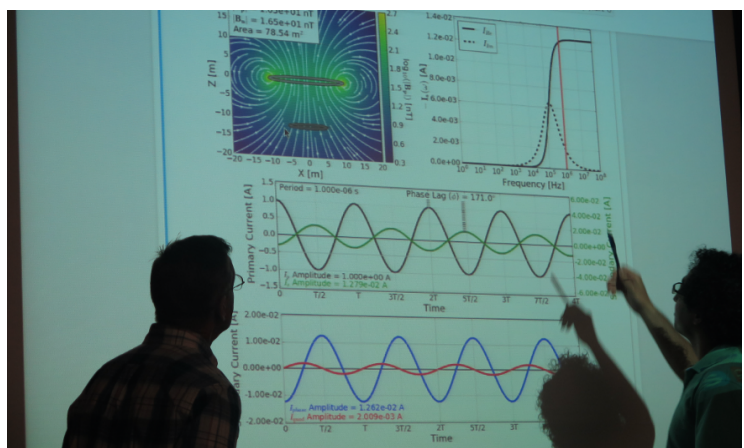

Figure 6. Using an app to explore the fundamentals of frequency domain electromagnetics at DISC 2017 at Kyoto University.

This was followed by grounded sources, whose understanding required a combination of material gained in DC and inductive sources. The energy requirements to excite large deep objects (base of salt, volcanoes, tectonic structures) are too big for controlled sources and hence natural sources are used. This gives rise to the MT (magnetotelluric) technique as well as ZTEM (Z-axis Tipper ElectroMagnetics). All of the EM methods outlined by that stage of the course made use of the quasi-static assumption, so that the energy diffuses into the earth. Although technically a wave, for frequencies less than $10^5$ Hz, the energy propagates only a fraction of a wavelength into the earth before it is severely attenuated, and so diffusive (quasi-static) approximations are sufficient. At high frequencies however, the energy travels as waves, and EM radargrams look much like seismic sections. As in seismic surveys, information about the substructure can be obtained by looking at plots of the data. This distinguishes it from the other EM techniques which require that the data be inverted to obtain physical property models that can reveal geologic information. The last technical topic addressed the fact that electrical resistivity can be frequency dependent for some materials. In practise this manifests itself as another set of charges that are induced in the earth; effectively the earth material acts like a capacitor. The IP (Induced Polarization) survey is designed to extract that signal and invert it to find chargeable material, such as in mineral bodies with sulfides.

Each section ended with one or more case histories that illustrated the use of the technique. By the end of the DISC we had collected many of these -- in large part thanks to external contributions -- and so, at each location, we were able to select case histories that were relevant to the local audience. Figure 7 illustrates some of these and shows the countries, problems, and the people who generated the formative material from which our case histories were built.

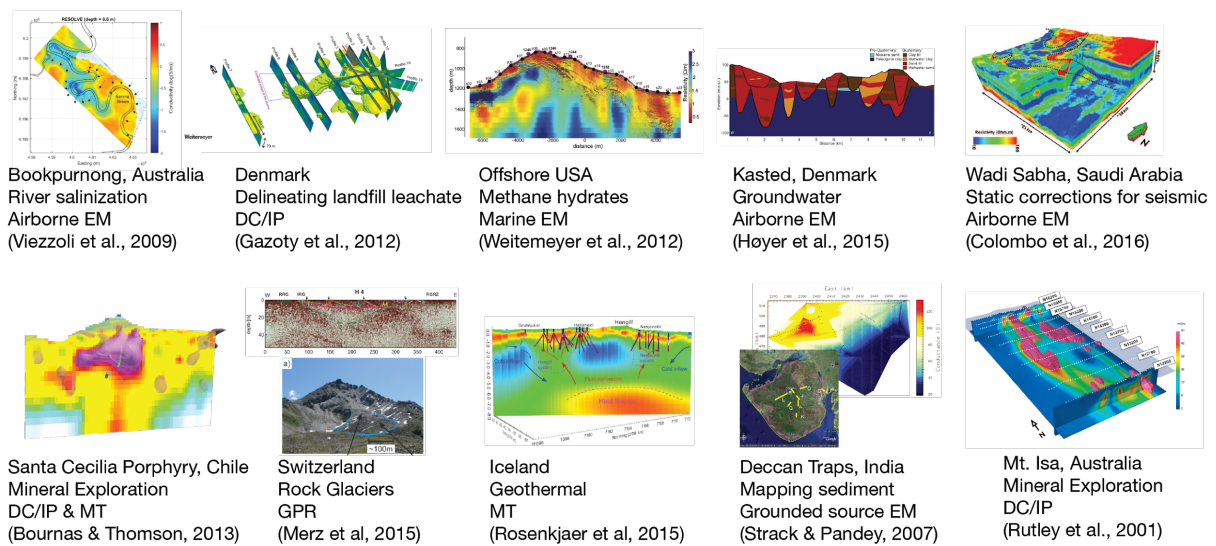

Figure 7: Sampling of Case Histories presented in DISC 2017.

The final section, "The Future" provided an opportunity to identify some of the important applications for EM and associated research problems that might occur.

## DISC Lab

Part of our motivation for DISC Lab was to use the DISC as an opportunity to have a two-way conversation. The course provided us with an opportunity to transmit information: introduce material, provide context, fundamentals, terminology. DISC Lab went the other way, giving participants a chance to discuss their work. Gaining insight about the problems people are working on, and the effectiveness (or not) of EM, was a primary goal.

The itinerary for the day was flexible and depended upon the audience. At some locations we spent the first part of the day providing additional course material and more advanced items, such as EM coupling removal for IP data. We then heard from attendees about their work in applying EM methods to problems of local interest. Generally these were 5-15 minute presentations followed by discussions. The talks (with permission from the presenter) were captured on YouTube, and links to the talks and associated slides are available from our blog (https://medium.com/disc2017). The variety of problems of interest in different countries was intriguing and insightful. At some locations hydrocarbons were the main interest, at others it was tectonics, minerals, geothermal energy, or environmental issues. However, at all locations water was a concern. This could be salt water intrusion, delineating aquifers, or using EM to build better hydrogeologic models.

The remaining time in the day was devoted to: (a) showing how to access the apps and show how they are developed in Jupyter notebooks; (b) an overview of SimPEG (Cockett et al., 2015; Heagy et al., 2017), a Python-based open-source library for numerical simulation and inversion; (c) tutorials for carrying out forward modelling and inversion. We used the recent publications in TLE (Kang et al., 2017) for this material. We ended by encouraging attendees to contribute to the information content on EM GeoSci.xyz and showing them how that can be done.

## What was accomplished?

We visited 20 countries and provided 24 presentations. Turn-outs were variable and ranged from a few 10's of people to over 100 (Figure 8a,b). India made the DISC into a special event, added a two-day conference, and arranged for people from across the country to attend. In Figure 8c, we show a map of where participants in India travelled from to attend the DISC. Further synoptic details about the location, date, attendees, and links to the course material, DISC lab and the blog site for each site can be found at: https://disc2017.geosci.xyz.

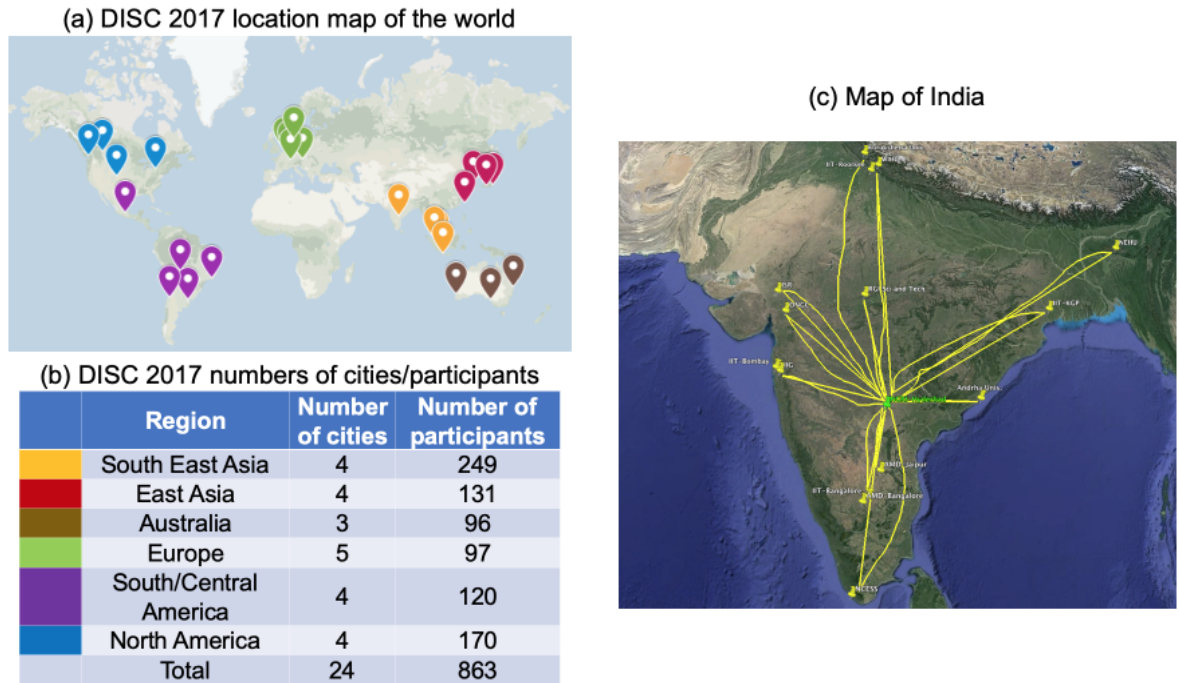

Figure 8. (a) locations of DISC 2017, (b) participation by region, and (c) locations where participants in India travelled from to attend.

EM was of interest in all countries visited but the applications were different and thus so were the surveys of interest. For instance: applications and countries ranking it as important include: (a) mineral exploration (Canada, Australia, Chile), (b) off-shore hydrocarbons (Netherlands, Malaysia, Brazil), (c) water issues (every country, although some were focussed upon finding the resource and others were concerned about contamination), (d) geothermal energy (most countries but particularly Chile and Germany) and (e) large-scale tectonics, particularly for earthquakes (Chile, India, Mexico City). These motivated the use of different EM surveys that had different depths of penetration, resolving power, and capabilities of being applied. Water issues made use of ground DC measurements and airborne EM, geothermal energy required MT to see deep structure, off-shore hydrocarbons required controlled-source EM (CSEM), environmental problems used galvanic and inductive source EM as well as ground penetrating radar, minerals applications used all of the techniques.

The audiences were diverse in their EM background and interests for application, and our challenge was to provide some information that would make an impact and promote the understanding and use of EM. We had much positive feedback about the course and content (and to pick a few of our favourites):

"Very good. I liked the focus on establishing a collaboration platform. The course has the potential of raising the bar and establishing common terminology." — Yusen Ley-Cooper, Geophysicist, Adelaide Australia

"The DISC for this year is NOT like a typical lecture in Japan. People are likely to be silent in many cases, but in this class, the audience members were very active in communication and in Q&A." — Koji Kashihara (Senior manager of R&D oil company)

"Really awesome. I could understand things I never could before. I loved how you showed [concepts] by images and videos [of] how the currents and magnetic fields are." — Daniela Montecinos, Santiago, Chile

It was gratifying for us that attendees said that they learned something new or that the course provided new insight about EM. To quote Dorris Lessing: "That is what learning is. You suddenly understand something you've understood all your life, but in a new way." A superb example of this was demonstrated in explaining the existence and interaction of the various EM waves that are of relevance in a marine EM survey. In particular the role of the air wave has been a source of confusion for many people (especially in early days of the technique). However, simulations and movies that showed fields and currents as a function of time (https://youtu.be/p2fjg5pcEQM) made the fundamental understanding about the process simple and intuitive. Figure 9 shows the four time snapshots of the electric fields illustrating the EM wave propagation in the marine EM survey setup.

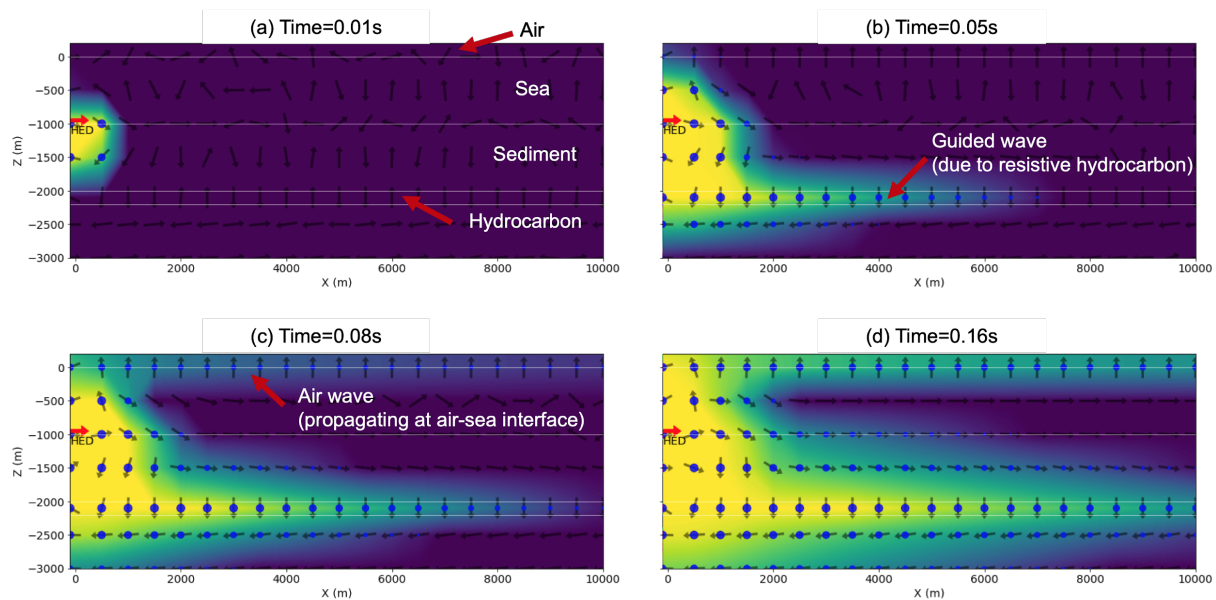

Figure 9. Propagation of the electric fields with the presence of the hydrocarbon at four times after the source current is turned off: (a) 0.01s, (b) 0.05s, (c) 0.08s, (d) 0.16s. The horizontal electric dipole (HED) source is located at the sea bottom. The black arrows indicate the direction of the electric field.

The ability to generate these movies, as well as all of the examples shown in the course, was made possible by the development of SimPEG, empymod (Werthmüller, 2017), and the use of Jupyter Notebooks. That has inspired others to proceed along these lines and it has spawned interactions between our research group at UBC and universities abroad.

The goal of DISC Lab was to discuss local problems for which EM could play a role and see how people were attacking those problems. Presentations spanned everything from instrument design, including a drone-based frequency domain system to locate vehicles buried in a landslide (Toyko, Japan), a time-domain EM system for groundwater (Brisbane, Australia) and a DC resistivity cart for soil characterization in vineyards (Vienna, Austria), to theoretical work on understanding the physical models of induced polarization (Mexico City, Mexico), to data processing and inversion for mineral exploration (Australia, Canada, Chile). Some presentations revealed new research avenues, and others raised questions that highlighted the need for resources on inverse theory -- inversion codes are powerful tools, but obtaining a meaningful result requires an understanding of optimization problems, how the algorithm operates, and the role of various hyperparameters in that algorithm. In many locations, we used some time on DISC Lab day to work through a hands-on tutorial of inversion using SimPEG.

In most cases, presenters allowed the material to be captured and these can be accessed through the blog. We were pleased with the engagement of participants and the breadth of discussions we had, especially in light of the fact that this was an entirely new format and it also required work on behalf of the participants.

## Post DISC 2017 and the future?

Our vision to foster the growth of a community of geoscientists who use EM remains undiminished and we continue to build upon the DISC 2017 materials and the broader collection of open-source resources in the GeoSci.xyz ecosystem. During the year 2017, we had a total of 26,744 unique visitors to EM GeoSci (Jan 1, 2017 - 2018). Usage has continued to grow and in the year 2019, over 100,000 unique visitors used the site. We have continued to build resources and courses from this material, including a course on Airborne Electromagnetics held as a part of the AEM 2018 conference in Denmark (https://courses.geosci.xyz/aem2018), and our resources have been adopted by others to serve as a textbook resource for graduate courses (Sun, 2018). Since all of the materials are licensed under a Creative Commons Attribution license, anyone is free to adapt and remix this content.

The SimPEG software, which was the foundation for the Jupyter Notebook "Apps", has continued to grow, and so too has the community that contributes to its development and support. The DISC and inversion tutorials that we ran on DISC Lab day motivated new developments within SimPEG, such as the ability to visualize charges in a DC resistivity simulation. Also, trying to explain computational steps within an educational context revealed confusing variable or method names; this prompted some refactoring to improve clarity. In a recent Geoscientists Without Borders project, we built upon our learnings from the DISC to create a course on groundwater exploration with geophysics in Myanmar (Fan et al., 2020). For that we also generated presentation materials and a series of Jupyter Notebooks that focussed on the basics of DC resistivity and inversion, as well as notebooks that local geoscientists could use to invert their field (Oldenburg et al., 2020). These resources are all available through GeoSci.xyz, and we hope they may serve other humanitarian geophysics efforts.

Expanding the scope of the DISC to a two-day event, developing a new model for a collaboratively created "textbook" resource, and distributing software as a part of the presentation materials was a challenging endeavor. However, the practices and tools we adopted to achieve this have changed how our GIF group carries out research, and illustrates the benefits to an open-source approach. Knowing that others may run your code, and potentially build upon it, encourages reproducible practices, including documenting software dependencies and testing, as well as "upstreaming" of fixes and new developments; this, goes beyond the usual practices of writing software that is sufficient to produce results for a scientific publication. Importantly, research results are achieved more quickly, and there is less duplication of efforts, when modular components can be reused. On GitHub, the SimPEG-research organization has been created to store Notebooks that are associated with publications and active projects so that results can be reproduced; some particular examples include (Astic, 2019; Kang, 2019; Heagy, 2019).

It is interesting to reflect how the increases in computational power, the advances made in forward modelling and inversion, and the advances in open source infrastructure have changed the fabric of research within our GIF group. Rather than being protective of intellectual property connected with inversion software we now realize that the value of technology resides not in the software itself but in those individuals who are making use of the codes. Moreover, because the development of software is most expediently done through collaboration with others, open source resources like EM GeoSci.xyz can play an important role. It is encouraging that the benefits of working in an open-source ecosystem are becoming widely accepted and numerous groups are contributing processing, inversion, geologic modelling, and 3D visualization tools.

In closing, the problems society faces, from resource management, to water scarcity and the broader impacts of climate change, are multidisciplinary challenges where EM can contribute. DISC 2017 was an opportunity to introduce participants to this ecosystem and a means for connecting the global EM community. Our hope is that the resources we developed, and the connections DISC 2017 enabled, help facilitate collaborations on problems where EM methods can play an important role.

# Acknowledgements

Thank you to all of the local hosts who arranged the course logistics and promoted participation in DISC 2017. Thanks to the SEG DISC program, and in particular Melissa Presson. We are also grateful for all of the contributors to EM GeoSci.xyz, including those who sent us Case Histories.